\documentclass[twocolumn,showpacs,preprintnumbers,amsmath,amssymb]{revtex4}

\usepackage{graphicx}
\usepackage{dcolumn}
\usepackage{bm}

\begin{document}


\title{Jahn-Teller instability in C$_6$H$_6^+$ and C$_6$H$_6^-$ revisited.}

\author{Vasili Perebeinos$^1$, Philip B. Allen$^2$, and Mark Pederson$^3$}
\affiliation{$^1$Department of Physics, Brookhaven National Laboratory, Upton,
New York 1973-5000,\\
$^2$ Department of Physics,
Department of Physics and Astronomy, State University of New York,
Stony Brook, NY 11794-3800 \\
$^3$Naval Research Laboratory, Complex Systems Theory Branch,
Washington, DC 20375-5345 }
\date{\today}

\begin{abstract}

The benzene cation (C$_6$H$_6^+$) has a doublet ($e_{1g}$) ground
state in hexagonal ring ($D_{6h}$) geometry.  Therefore a
Jahn-Teller (JT) distortion will lower the energy.
The present theoretical study
yields a model H\"{u}ckel-type Hamiltonian that includes the JT coupling of the
$e_{1g}$ electronic ground state with the {\it two} $e_{2g}$
vibrational modes: in-plane ring-bending and C-C bond-stretching.
We obtain the JT couplings from density functional theory (DFT),
which gives a JT energy lowering of 970 cm$^{-1}$ in agreement
with previous quantum chemistry calculations.
We find a non-adiabatic solution for vibrational spectra and
predict frequencies shifts of both the benzene cation and anion,
and give a reinterpretation of the available experimental data.
\end{abstract}


\maketitle

\section{\label{sec1}Introduction}

A molecular system with a degenerate electronic ground state will
spontaneously lower its symmetry to lift the degeneracy. This is
the Jahn-Teller (JT) \cite{Jahn} effect, and is common in
molecules \cite{Bersuker}. The C$_6$H$_6$ (benzene) molecule has a
long history of study \cite{Kistiakowsky,Sponer}.
The benzene cation represents a paradigm of the dynamical JT
effect \cite{Longuet}.
With full D$_{6h}$ symmetry,
the highest occupied molecular orbital (HOMO) $e_{1g}$ is doubly
degenerate, as is the lowest unoccupied molecular orbital (LUMO)
$e_{2u}$. Hence both the cation and anion are JT unstable and
prefer lower symmetry.  In the lowest (linear) approximation, the
Born-Oppenheimer surface has a degenerate loop in configuration
space surrounding the D$_{6h}$ ground state.  Everywhere on this
loop, the HOMO and LUMO states are split by a constant JT gap. The
most symmetric points on this loop represent simple D$_{2h}$
distortions.  These distortions occur in three equivalent
``acute'' and three equivalent ``obtuse'' forms (see the inset to
Fig. 2). Higher order corrections lift the degeneracy on the loop,
but zero-point energy exceeds the barrier height, creating a
dynamic rather than a static JT distortion.

{\it Ab initio} approaches \cite{Engstrom,Koppel,Raghavachari}
including density functional theory (DFT) \cite{Muller,Yoshizawa}
have been used to predict the ground state geometry and the JT
energy lowering in the benzene cation.
Quantum chemistry methods have been used to analyze vibrational spectra of
the benzene cation \cite{Eiding,Pulay,Lipari}.
A high resolution
zero-kinetic-energy (ZEKE) photoelectron study by Linder {\it et
al.} \cite{Linder} found values of the frequency shift of the
in-plane ring-bending $e_{2g}$ vibrational mode.  From their
analysis, they deduced the JT energy lowering to be 208 cm$^{-1}$.
They also concluded that the cation has a global minimum for the acute
D$_{2h}$ geometry, with a local minimum for the obtuse geometry only 8
cm$^{-1}$ higher in energy.  {\it Ab initio} calculations
\cite{Raghavachari} suggest 4-5 times greater energy gain. The
present study reconciles this discrepancy.  We provide a
microscopic model Hamiltonian for the $\pi$-electron system, which
includes two independent $e_{2g}$ modes of JT coupling.  We derive
all parameters of this model from DFT calculations
and find vibrational spectra for both the cation and
anion using a non-adiabatic approximation.
Our results for the cation permit a reinterpretation of the
ZEKE spectra, allowing a larger JT energy lowering, which agrees
with both our DFT and previous theory.

\section{\label{sec2}Model Hamiltonian}

The starting point is a H\"{u}ckel-type Hamiltonian for the
non-bonding
$\pi$ electrons, with a first-neighbor hopping integral
$t_1=-<i|{\cal H}|i\pm1>$ between adjacent $p_z$ atomic orbitals
on the six C atoms. In Slater-Koster notation \cite{Slater},
$-t_1$ is called the $(pp\pi)$ two-site integral. In the D$_{6h}$
point group, all six $t_1$ integrals are the same. The
6-dimensional electronic Hilbert space has single-particle states
with $a_{2u}$, $e_{1g}$, $e_{2u}$, $b_{1g}$ symmetries, which have
energies $-2t_1,-t_1,t_1,2t_1$.

Hopping matrix elements $t$ decrease with distance between atoms.
For small atomic displacements, $t$ depends linearly on the C-C
bond length, $t_1(\delta R_{i,i+1})=t_1^0-g_1\delta R_{i,i+1}$.
The electron-phonon coupling constant $g_1$ is the same as in the
Su-Schriefer-Heeger model\cite{Su} for polyacetylene.  Next we
introduce second neighbor hopping
$t_2=<i-1|H|i+1>=t_2^0+g_2\delta\alpha_i$ to include a dependence
on the change $\delta\alpha_i$ of the C$_{i-1}$-C$_{i}$-C$_{i+1}$
bond angle $\alpha_i$ from 120$^{\circ}$.   The constant term
$t_2^0$ does not lift the degeneracy of the $e$ levels.  Therefore
we put $t_2^0=0$ to keep the Hamiltonian as simple as possible. We
use harmonic restoring forces $K_1, K_2$ for the bond stretching
and angle bending vibrations.  The total Hamiltonian ${\cal
H}={\cal H}_{\rm el}+{\cal H}_{\rm vib}$ has the form
\begin{eqnarray}
{\cal H}_{\rm el}&=&\sum_{i=1}^6\left[-t_1^0+g_1\delta R_{i,i+1}
|i><i+1|\right. \nonumber\\ &&\left. +g_2 \delta \alpha_{i}
|i-1><i+1|+h.c. \right] \nonumber\\ {\cal H}_{\rm
vib}&=&\sum_{i=1}^6(P^2_{i}/2M +K_1\delta R_{i,i+1}^2/2+K_2\delta
\alpha_{i}^2/2). \label{hmodel}
\end{eqnarray}

\begin{figure}
\includegraphics[height=2.05in,width=2.85in,angle=0]{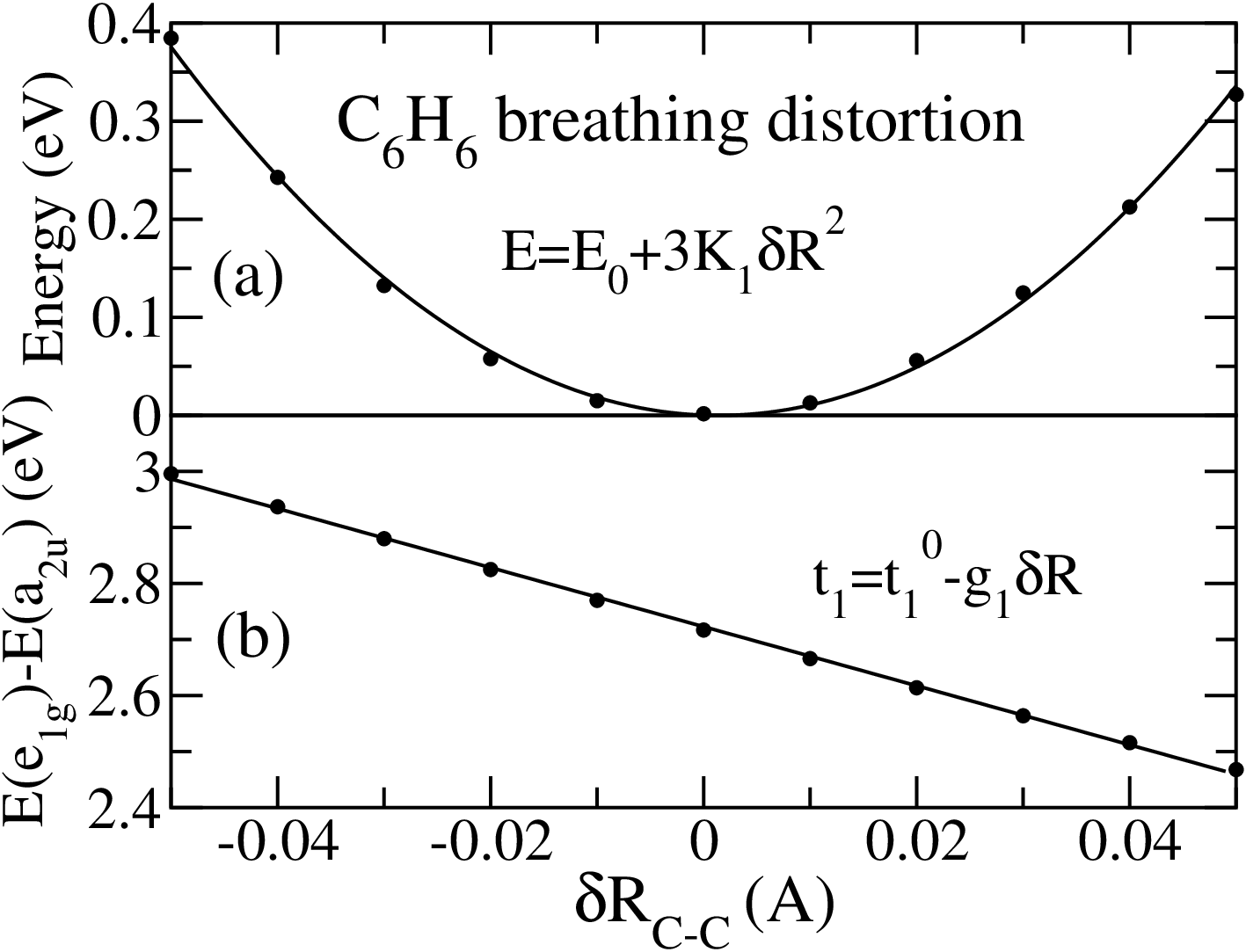}
\caption{\label{fig1}
(a) DFT total energy for neutral benzene with a symmetric
a$_{1g}$ distortion around the equilibrium C-C separation (closed
circles).  (b) Eigenvalue difference for the two lowest $\pi$
states (closed circles).  The C-H bond lengths are held fixed for
both calculations. The solid curves are best fits using the model
Hamiltonian Eq. (\protect\ref{hmodel}) with $g_1$ and $K_1$
adjusted.}
\end{figure}

The coupling constants $g_{1,2}$ and spring constants $K_{1,2}$
are calculated from DFT using the program NRLMOL\cite{Mark} with a
Gaussian basis set and the Perdew-Burke-Ernzerhof
exchange-correlation potential \cite{Perdew}. This yields as the
optimal geometry of neutral benzene in D$_{6h}$ symmetry, the C-C
and C-H bond lengths of $R_0$=1.398 \AA~ and 1.08 \AA~ respectively.
The energy difference between the HOMO $e_{1g}$ and $a_{2u}$
orbitals defines the hopping integral $t_1^0=2.72$ eV.  We fix the
C-H bond length and introduce a symmetric $a_{1g}$ breathing
distortion, which alters the C-C bond lengths.  The results, shown
on Fig. (\ref{fig1}), can be fitted to the predictions of the
model Hamiltonian (\ref{hmodel}) $E=E_0+3K_1\delta R^2$ (for the
total energy) and $\Delta E=t_1=t_1^0-g_1\delta R$ for the
$E(e_{1g})-E(a_{2u})$ eigenvalue difference.  The results are
$g_1=5.27$ eV/$\AA~$ and $K_1=47.4$ eV/$\AA^2$.

\begin{figure}
\includegraphics[height=1.99in,width=2.85in,angle=0]{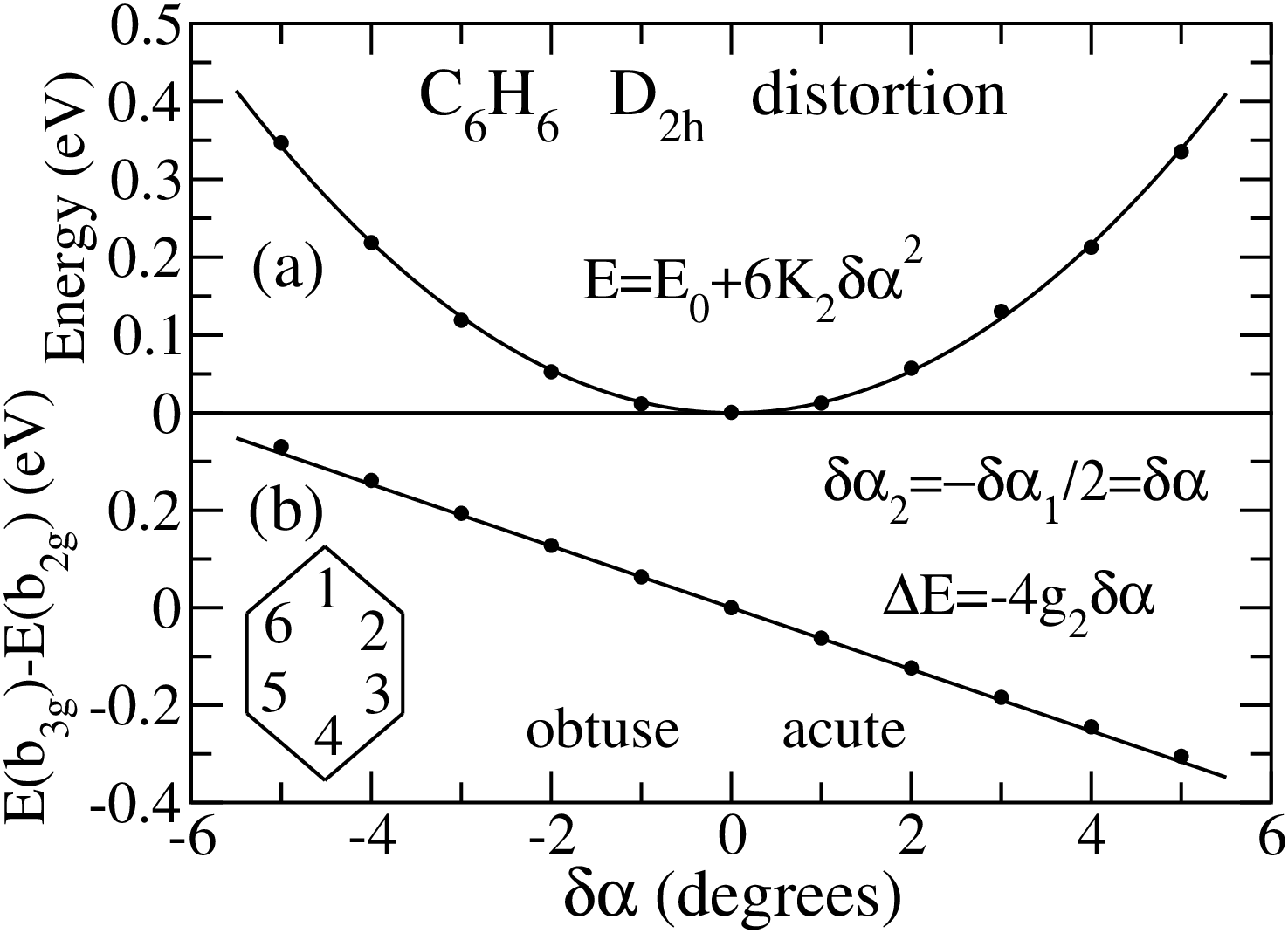}
\caption{\label{fig2}
(a) DFT total energies and (b) splitting of the doublet
HOMO eigenvalues of neutral benzene with a D$_{2h}$ distortion and
fixed C-C bond lengths as shown in the inset.  The solid curves
are best fits using  the model Hamiltonian Eq.
(\protect\ref{hmodel}) to fix the parameters $g_2$ and $K_2$.}
\end{figure}

Second neighbor integrals are found from DFT energies for distortions
with D$_{2h}$ symmetry.   We fixed the C-C bond lengths and
varied the bond angle $\delta \alpha$ as shown in the inset
to Fig. (\ref{fig2}). Specifically, angles 1 and 4 were decreased by
2$\alpha$, while the remaining four angles were increased by
$\alpha$.  The hydrogen atoms were fixed along the bisectors
of the C-C-C angles.  The results were fitted to the predictions of
the Hamiltonian (\ref{hmodel})
$E=E_0+6K_2\delta \alpha^2$ for the total energy, and
$4g_2\delta\alpha$ for the splitting
of doubly degenerate $e_{1g}$ HOMO states into $b_{2g}$ and
$b_{3g}$ singlets.  The results are $K_2=7.45$ eV/rad$^2$
and $g_2=0.91$ eV/rad.

\section{\label{sec3}Adiabatic Solution}

For fixed atomic coordinates, the adiabatic potential energy
surface (APES) can easily be calculated using Eq. (\ref{hmodel}).
For the D$_{2h}$ distortion shown in the inset of Fig.
(\ref{fig2}), the energy depends on three variables $\delta
R_{1}=\delta R_{12}$, $\delta R_{2}=\delta R_{23}$ and $\delta
\alpha_2=-\delta \alpha_1/2=\delta \alpha$.  The electronic
contribution is the sum of the occupied eigenvalues of Eq.
(\ref{hmodel}).
The energy is not analytic at the point ($\delta R_{1}, \delta
R_{2}, \delta \alpha$)=(0,0,0), but has a cusp, characteristic of
the first-order JT splitting.
Charged benzene lowers its energy by $\delta E_1+\delta E_2
+\delta E_3$ corresponding to atomic relaxation {\it via} three
types of vibrational modes. The first is the symmetric $A_{1g}$
distortion of D$_{6h}$ symmetry, $\delta R_1=\delta R_2=\delta
R_0$ and $\delta\alpha=0$.
The second and the third are two $E_{2g}$ vibrational modes:
the ring-bending and C-C bond-length distortions.
It is more convenient to measure
distortions $\delta R'$ around the relaxed breathing positions
$R=R_0+\delta R_0=1.416$ \AA.
The distorted geometry of benzene ions and corresponding energy
gains are shown in Table \ref{table1}.

\begin{table}
\begin{ruledtabular}
\begin{tabular}{rrrrr}
  & Model & & DFT & \\
\hline
   $\delta R_0 \ (\AA)$ & 0.018 &(0.018) & 0.014 & (0.022) \\
   $\delta E_1$ (eV) & -0.049 &(-0.049)  & -0.028 & (-0.046)\\
\hline
  & acute & obtuse & acute  & obtuse   \\
\hline
   $\delta\alpha_1=-2\delta\alpha$
  & 2.36$^{\circ}$ & -2.38$^{\circ}$ & 1.65$^{\circ}$ & -1.73$^{\circ}$ \\
   $\delta E_2$ (eV) & -0.019 & -0.019 &  & \\
\hline
   $\delta R_1^{\prime}$ \ ($\AA$)
  & 0.020 & -0.020 & 0.020 & -0.020 \\
   $\delta R_2^{\prime}$ \ ($\AA$)
  & -0.040 & 0.040 & -0.037 & 0.042 \\
   $\delta E_3$ (eV) & -0.105 & -0.108 &  & \\
\hline
 $\Delta E_{\rm JT}^{\rm tot}$ \ (eV)
  & -0.124 & -0.127 & -0.123 & -0.120 \\
  $2\Delta_{\rm JT}$ \ (eV)
  & 0.497 & 0.508 & 0.500 & 0.481 \\
\end{tabular}
\end{ruledtabular}
\caption{\label{table1} The geometry and energies of the model Hamiltonian and
DFT solutions for the benzene cation.
The breathing $a_{1g}$ mode relaxation for
the anion are shown in brackets.}
\end{table}

The geometry and energies of the DFT
solution (Table \ref{table1})
are reasonably well reproduced by the model Hamiltonian.
The energy difference (3.7 meV) found between the
acute or obtuse distortions is sensitive to the choice of
exchange-correlation potential and smaller than the errors of
calculations \cite{Muller}.
The DFT energy lowering and carbon atomic distortions are affected
by hydrogen distortions as well, which are left out completely in
Eq. (\ref{hmodel}). We did constrained geometry optimization with
hydrogens forced to follow carbons so that C-H bond divides C-C-C
angle by half. The minimum energy in the constrained calculations was
only 4 meV higher in energy than in the fully optimized geometry for both
acute and obtuse cations.
The main conclusion is that the C-C
bond-length distortions contribute the most to the JT energy
lowering and degenerate level splitting in both the benzene cation
and anion.

\section{\label{sec4}Vibrational Model Hamiltonian}

JT-coupled vibrational modes change their frequencies on a charged
molecule. The observable frequency shifts serve as a measure of
the JT energy lowering. We solve the model Hamiltonian for the
vibrational spectra for both cation and anion.  For the case of
the cation, the Hamiltonian (\ref{hmodel}) is projected on the two
$e_{1g}$ HOMO states with symmetries $b_{3g}$ and $b_{2g}$ in the
$D_{2h}$ point group,
\begin{eqnarray}
&&|b_{3g}>=\frac{1}{\sqrt{12}}\left(2|1>+|2>-|3>-2|4>-|5>+|6>\right)
\nonumber\\
&&|b_{2g}>=\frac{1}{2}\left(|2>+|3>-|5>-|6>\right)
\label{b2g}
\end{eqnarray}
The Hamiltonian takes form
$U_1HU_1^{\dagger}=H_0\hat{I}+H_z\sigma_z+H_x\sigma_x$,
where $U_1=\left(|b_{3g}>, |b_{2g}>\right)$, and
$\sigma_{\beta}$ are the Pauli matrices in the
$\{b_{3g},b_{2g}\}$ subspace,
\begin{eqnarray}
{\cal H}_0=&&\sum_{i=1}^6\left(P^2_{i}/2M
+K_1\delta R^{'2}_{i,i+1}/2+K_2\delta \alpha_{i}^2/2\right)
\nonumber\\
{\cal H}_z=&&\frac{g_1}{6}(\delta R'_1+\delta R'_3+\delta R'_4+
\delta R'_6-2\delta R'_2-2\delta R'_5)
\nonumber\\
+&&\frac{g_2}{2}(\delta\alpha_1+\delta\alpha_4)
\nonumber\\
{\cal H}_x=&&\frac{g_1}{\sqrt{12}}(\delta R'_1-\delta R'_3+\delta R'_4-
\delta R'_6)
\nonumber\\
+&&\frac{g_2}{\sqrt{12}}(\delta\alpha_2-\delta\alpha_3+
\delta\alpha_5-\delta\alpha_6)
\label{Hpr}
\end{eqnarray}

In adiabatic approximation, the JT orbital splitting is
$2\Delta=2g_1^2/3K_1+2g_2^2/3K_2$ and the JT energy lowering is
$E_{\rm JT}=-\Delta/2$. To solve the problem beyond the adiabatic
approximation we have to quantize vibrational motions. We rewrite
the Hamiltonian (\ref{Hpr}) in Cartesian coordinates $(\delta x_i,
\delta y_i)$. The kinetic energy term is diagonal. Choosing as the
unit of displacement $R$=1.416 \AA (the benzene ring radius after
$a_{1g}$ mode relaxation), and as the energy unit $K_1R^2$, the
potential energy of the Hamiltonian $H_0$ (Eq. (\ref{Hpr}))
depends on a single parameter $K_2/K_1R^2=\Lambda$.

After diagonalizing $H_0$ we get twelve normal modes whose
amplitudes we designate $\Theta_{1..12}$. The potential energy is
independent of the motion of the center of mass ($X=\sum
x_i$,$Y=\sum y_i$) and the ring rotation ($\alpha=\sum \alpha_i$).
This forces three normal modes to have frequencies zero. Since we
neglect C-H stretching motions, the mass M is $M_C+M_H$. Choosing
$\Lambda=0.0784$ the nine remaining normal modes are: the
degenerate ring bending $E_{2g}$ mode at $\omega_1=581$ cm$^{-1}$
with some admixture of the C-C bond length alteration, the pure
ring bending $B_{1u}$ mode at 962 cm$^{-1}$, the breathing
$A_{1g}$ mode at 992 cm$^{-1}$, the mostly C-C
bond-length-alteration doublets $E_{1u}$ and $E_{2g}$ at 1262
cm$^{-1}$ and $\omega_2=1644$ cm$^{-1}$ and the pure C-C
bond-stretching mode $B_{2u}$ at 1718 cm$^{-1}$. It is interesting
to compare the model Hamiltonian normal mode spectrum with the DFT
NRLMOL calculations. There are 21 in-plane vibrational modes
$\Gamma=2A_{1g}+A_{2g}+4E_{2g}+2B_{1u}+2B_{2u}+3E_{1u}$.  Six of
them are high frequency C-H modes $A_{1g}+B_{1u}+E_{1u}+E_{2g}$
around $3100$ cm$^{-1}$.  From the remaining 15 vibrations we
identify the ring-bending $E_{2g}$ mode at $600$ cm$^{-1}$, the
breathing $A_{1g}$ at $992$ cm$^{-1}$, and the bond-stretching
$E_{2g}$ at $1589$ cm$^{-1}$.

The off diagonal part $H_x$ of the Hamiltonian (\ref{Hpr})
depends only on the two JT-active $E_{2g}$ vibrational normal modes.
Let us label the amplitudes of these modes by
($\Theta_{1a}, \Theta_{1b}$) and ($\Theta_{2a}, \Theta_{2b}$) and
denote their frequencies by $\omega_1$ (ring bending) and
$\omega_2$ (bond-stretching) as above. We also use the coordinates
$\Theta_{1}$, $\Theta_{2}$ to denote the total displacements
$\Theta_{1}^2=\Theta_{1a}^2+\Theta_{1b}^2$ etc.
Then Eq. (\ref{Hpr}) becomes:
\begin{eqnarray}
{\cal H}_0=&&\frac{M}{2}\left(\dot{\Theta}_{1a}^2+\dot{\Theta}_{1b}^2+
\omega_1^2\Theta_1^2\right)\nonumber\\
+&&\frac{M}{2}\left(\dot{\Theta}_{2a}^2+\dot{\Theta}_{2b}^2+
\omega_2^2\Theta_2^2\right)\nonumber\\
{\cal
H}_z=&&-G_1\Theta_1\cos(\beta_1)-G_2\Theta_2\cos(\beta_2)
\nonumber\\ {\cal
H}_x=&&-G_1\Theta_1\sin(\beta_1)-G_2\Theta_2\sin(\beta_2),
\label{Hp2}
\end{eqnarray}
where the angles $\beta_1$, $\beta_2$ contain arbitrary additive constants
which are the choices of the orientation angles of the orthogonal eigenvectors
in the two-dimensional $E_g$ subspaces of the two modes.
Note that the adiabatic electronic eigenvalues which diagonalize
${\cal H}_z\sigma_z+{\cal H}_x\sigma_x$ (in the approximation that
($\Theta_{1a},..,\Theta_{2b}$) are classical variables) are $\pm\Delta_{JT}$
where
\begin{eqnarray}
\Delta_{JT}^2=G_1^2\Theta_{1}^2+G_2^2\Theta_{2}^2+2G_1G_2\Theta_{1}\Theta_{2}
\cos(\beta_1-\beta_2) \label{DJT}
\end{eqnarray}
This does not depend separately on the angles $\beta_1$, $\beta_2$.
The modified electron-phonon coupling constants $G_1$, $G_2$ depend on the
JT couplings $g_1$, $g_2$ and on the $E_g$ normal mode eigenstates.
For our case ($\Lambda=0.0784$) we find
\begin{eqnarray}
G_1&=&T_1g_1/6+T_2g_2/2R=1.3 \ \ {\rm eV/\AA~}
\nonumber\\
G_2&=&T_3g_1/6-T_4 g_2/2R=4.14 \ \  {\rm eV/\AA~}
\label{G12}
\end{eqnarray}

Let us consider further the adiabatic eigenstates of ${\cal
H}={\cal H}_z\sigma_z+{\cal H}_x\sigma_x$. What happens when we
change the angle
$\delta\beta_1=\delta\tan^{-1}\Theta_{1b}/\Theta_{1a}$? The
rotation matrix
$R=\cos(\alpha/2)\hat{\sigma}_z+i\sin(\alpha/2)\hat{\sigma}_x$ has
the property that $R{\cal H}R^{\dagger}$ gives a new matrix ${\cal
H'}$ with angles $\beta_1\rightarrow\beta_1+\alpha$ and
$\beta_2\rightarrow\beta_2+\alpha$. The energies $\pm\Delta_{JT}$
are unchanged and the eigenfunctions are rotated in ($b_{3g}$,
$b_{2g}$) space by $\alpha/2$. This type of change of distortion
pattern is called a ``pseudo-rotation''. A full pseudo-rotation by
$\alpha=2\pi$ restores the molecular geometry and changes the sign
of the electronic eigenstate. This sign change under adiabatic
distortion is the Berry phase, and needs to be correctly
incorporated when finding the system's total wavefunction in
adiabatic approximation. In particular it leads to fractional
quanta of pseudorotational angular momentum. Our method will find
total wavefunctions without making the adiabatic approximation and
will therefore yield correct answers without explicit mention of
Berry phase.

\section{\label{sec5}Non adiabatic Solution}

The quantized vibrational Hamiltonian (\ref{Hp2}) depends on three
parameters $\omega_1/\omega_2=0.353$ and the two electron-phonon
coupling constants $\kappa_1=(\Delta_1/\hbar\omega_1)^{1/2}=1.18$,
$\kappa_2=(\Delta_2/\hbar\omega_2)^{1/2}=0.8$, where
$\Delta_1=G^2_1/M\omega_1^2=817$cm$^{-1}$ and
$\Delta_2=G^2_2/M\omega_2^2=1052$cm$^{-1}$  are  twice the JT
energy gain due to the ring-bending and the bond stretching modes
correspondingly. Similar values of $\kappa_1=1.06$ and
$\kappa_2=0.817$ were obtained by Eiding {\it et al.}
\cite{Eiding} using {\it ab initio} many body techniques.

To solve Eq. (\ref{Hp2}) for the vibrational spectrum, assume for
the moment a zero coupling to the bond stretching vibrational mode
$G_2=0$. Then Eq. (\ref{Hp2}) depends on a single parameter
$\kappa_1$. The classical solution of Longuet-Higgins {\it et al.}
\cite{Longuet} uses polar coordinates, which simplifies equations.
Instead we seek the solution of Eq. (\ref{Hp2}) in the form
\begin{eqnarray}
|\Psi>=\sum_{n_1=0}^N\sum_{n_2=0}^{N-n_1}
&&\left[A_{n_1,n_2}\frac{(c_a^{\dagger})^{n_1}}
{\sqrt{n_1!}}\frac{(c_b^{\dagger})^{n_2}}{\sqrt{n_2!}}|0>|b_{3g}>
\right. \nonumber\\ +&& \left.
B_{n_1,n_2}\frac{(c_a^{\dagger})^{n_1}}
{\sqrt{n_1!}}\frac{(c_b^{\dagger})^{n_2}}{\sqrt{n_2!}}|0>|b_{2g}>
\right] \label{Hp3}
\end{eqnarray}
%
Note that we are not making the adiabatic approximation where the
electronic wavefunction $\Psi_{\rm el}$ is strictly held in the
lower of the two Jahn-Teller split levels. In adiabatic
approximation the double valuedness of  $\Psi_{\rm el}$ under
rotation by $\beta=2\pi$ is compensated by a restriction on the
vibrational wavefunctions. This effect is automatically handled in
our non-adiabatic treatment. The operators $c_{a}^{\dagger}$,
$c_{b}^{\dagger}$ create harmonic oscillator states in the
two-dimensional manifold of $e_g$ vibrations, with their origin at
the symmetric ($D_{6h}$) benzene ion coordinates. Although
different from the basis of Longuet-Higgins {\it et al.}
\cite{Longuet}, both basis sets are complete and converge rapidly
to the same solution.  The solution is found by exact
diagonalization of Eq. (\ref{Hp2}) for $N=10$. The eigenvalues
coincides with those of Linder et. al. \cite{Linder} for the case
of zero quadratic coupling.

\begin{figure}
\includegraphics[height=1.99in,width=2.63in,angle=0]{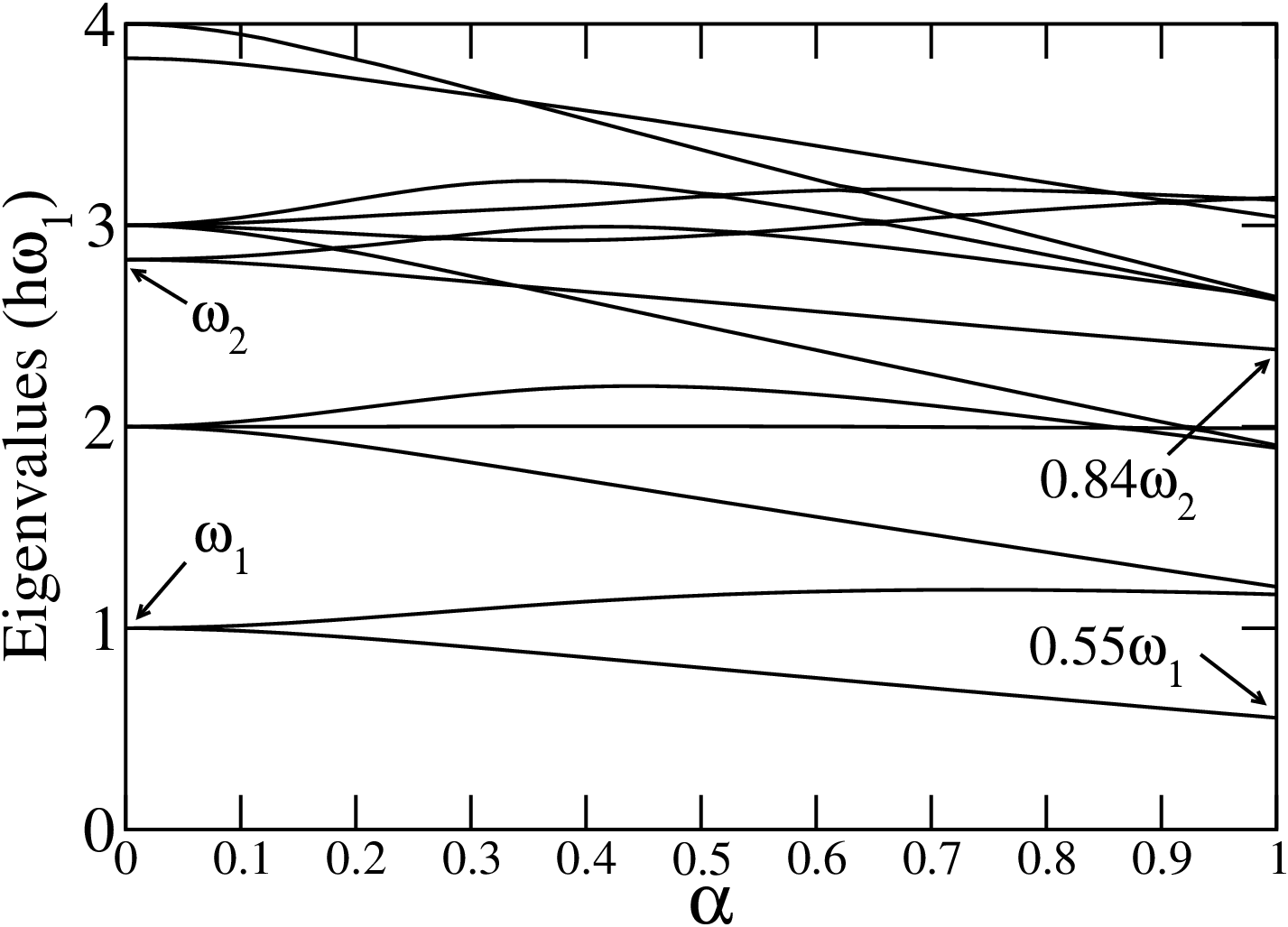}
\caption{\label{fig3}
Vibrational spectrum of the Jahn-Teller unstable ion {\it
versus} parameter $\alpha$, where coupling constants
$\kappa_1=1.18*\alpha$ and $\kappa_2=0.8*\alpha$.
For the cation, the coupling
constant for the ring bending mode is $\kappa_1=1.18$ and
coupling to the bond stretching mode is $\kappa_2=0.8$.
Therefore for $\alpha=1$
the neutral molecule vibrational frequency $\omega_1=581$
cm$^{-1}$ becomes 0.55$\omega_1$ and
$\omega_2=1644$ cm$^{-1}$ reduces to 0.84$\omega_2$
for the cation.}
\end{figure}

The vibrational frequencies on the neutral molecule $\omega_1=581$
cm$^{-1}$ and $\omega_2=1644$ cm$^{-1}$ become (for the benzene
cation) 0.54$\omega_1$ and
0.7$\omega_2$ in the uncoupled-mode approximation.
To take into
account the interaction between the two $E_{2g}$ vibrational modes
(when both $G_1\neq 0$ and $G_2\neq 0$) one has to use a trial
wavefunction similar to Eq. (\ref{Hp3}), but including all four
vibrational modes.  The size of the matrix grows as $2\times
C_{N+4}^4\approx (N+2.5)^4/12$. Convergence was reached for
$N=10$. The lowest eigenvalues as a function of
parameter $\alpha$ are shown on Fig. \ref{fig3}.
The coupling constants are proportional to $\alpha$ and at $\alpha=1$
they become $\kappa_1=1.18$ and $\kappa_2=0.8$ .
The frequency shift of the ring bending mode $\omega_1\rightarrow
0.55\omega_1$ is essentially the same as in the noninteracting
case. The interaction of the bond stretching mode with the ring
bending overtones results in the level repulsion, such that
$\omega_2\rightarrow0.84\omega_2$ for $\kappa_2=0.8$. The
predicted vibrational spectrum corresponds to $\alpha=1$ on Fig.
\ref{fig3}.

On the anion, a $D_{2h}$ distortion splits the $e_{2u}$ LUMO into
two $b_{1u}$ and $a_u$ orbitals, which can be obtained from Eq.
(\ref{b2g}) by changing the sign in front of the $|2>$, $|4>$,
$|6>$ molecular orbitals.
%
%
Projecting the Hamiltonian Eq. (\ref{hmodel}) into the
$\{b_{1u},a_u\}$ manifold leads to the same Hamiltonian as
(\ref{Hp2}) with a sign change of coupling constant
$g_1\rightarrow -g_1$.  The vibrational excitation spectra on the
anion $C_6H_6^-$ is described by Eq. (\ref{Hp2}), with coupling
constants $G'_1=-T_1 g_1/6+T_2 g_2/2R_0=0.17$ eV/\AA~ and
$G'_2=-T_3 g_1/6-T_4 g_2/2R_0=-5.48$ eV/\AA.~ Since the $H_0$ part
of the projected Hamiltonian is the same as for the cation, the
values for $T_{1..4}$ are unchanged. The JT energy lowering on the
anion is the same as for the cation
$G'^2_1/M\omega_1^2+G'^2_2/M\omega_2^2=0.234$ eV, but the
effective couplings to the ring bending and C-C bond stretching
modes are now different; $\kappa'_1=0.16$ and $\kappa'_2=1.06$
respectively.  This means that the frequency of the ring bending
$E_{2g}$ mode is shifted by less then 2\% from the neutral benzene
value, whereas the bond-stretching $E_{2g}$ mode becomes 59\% of
the neutral molecule value 1644 cm$^{-1}$.

\section{\label{sec6}Summary}

Using the model Hamiltonian Eq. (\ref{hmodel}) with DFT-calculated
parameters, we predict the JT stabilization energy of 944
cm$^{-1}$ for both the benzene cation and anion.  The
bond-stretching mode 1644 cm$^{-1}$ on neutral benzene experiences
a large shift on the anion and becomes 970 cm$^{-1}$, which is
very close in energy to the $a_{1g}$ breathing mode
$992$cm$^{-1}$.  On the other hand, the ring-bending mode is
essentially unaffected.  To the contrary, on the cation both
$E_{2g}$ JT active modes experience the sizable shifts by 45\% and
16\% from their neutral benzene values.  Experimentally
\cite{Linder,Goode} the ring-bending mode on the benzene cation
was measured to reduce from $536$cm$^{-1}$ to $350$cm$^{-1}$ (or
by 35\%).  This measurements led Linder {\it et al.} \cite{Linder}
to deduce a JT energy stabilization of 208 cm$^{-1}$.  A more
detailed analysis, including quadratic JT coupling, led to the
conclusion that the acute $D_{2h}$ distorted geometry is the
global minimum, lower by 8 cm$^{-1}$ than the obtuse local
minimum.  Our Hamiltonian (\ref{Hp2}) predicts a larger relative
shift of the ring bending mode by 10\% with respect to the
experimental shift. The amount of the shift is sensitive to the
choice of normal modes determined by the term ${\cal H}_0$ (Eq.
\ref{Hp2}), a slightly oversimplified model. The main result of
our new solution is that there are {\bf two} contributions to the
JT energy lowering.  To determine the JT energy lowering on the
cation {\bf two} vibrational $E_{2g}$ frequencies  have to be
measured.  The model Hamiltonian (\ref{hmodel}) contains quadratic
couplings, which predict an acute global minimum, while the
accuracy of the true DFT answer is insufficient to make such a
prediction \cite{Muller}.

The benzene anion is not stable, which makes it difficult to
measure the vibrational spectrum.  Electron transmission
techniques \cite{Burrow} give an opportunity to measure the
vibrational sidebands in the resonant electron cross section at
negative electron affinity energy 1.1 eV \cite{Burrow}.  The
spectrum shows vibrational peaks separated by 123 meV (or 992
cm$^{-1}$), and were attributed to $a_{1g}$ vibrational quanta.
The higher frequency JT-active $e_{2g}$ mode has not been
resolved.  Our results, however, suggest that since both breathing
and C-C bond stretching modes are coupled to the LUMO, a mixture
of the two is expected to be present in the spectrum.  Since the
$e_{2g}$ modes becomes very close in energy to the breathing
$a_{1g}$ mode, it is difficult to distinguish experimentally
between the two.

In conclusion we derived a new model Hamiltonian for the JT active
benzene cation and anion, and obtained all parameters from DFT
calculations.  This model Hamiltonian predicts a JT stabilization
energy of $970$ cm$^{-1}$.  This value, similar to other
first-principles results, is 4.5 times larger than the value
deduced experimentally by Linder {\it et al.} \cite{Linder}.
However, our model gives the result that {\bf two} $E_{2g}$ modes
are JT-active, coupling to the electronic HOMO and LUMO states.
Only one of them was measured \cite{Linder}, and our results are
reasonably consistent with that measurement.  However, both modes
are needed to obtain the JT stabilization energy from
spectroscopic data.  We conclude that there is no contradiction
between theory and experiment, and predict the relevant frequency
shifts that can be used to test theory more completely.

\begin{acknowledgments}
We are grateful to Prof. Philip M. Johnson for useful discussions.
This work was supported in part by DOE  Grant No.\ DE-AC-02-98CH10886,
and in part by NSF Grant No. DMR-0089492.
\end{acknowledgments}

\end{document}